\documentclass[12pt]{JHEP}

\usepackage{amsmath}
\usepackage{amssymb}

\def\sla#1{\raise.15ex\hbox{/}\kern-.57em #1}
\def\slas#1{\raise.15ex\hbox{/}\kern-.72em #1}

\title{Relating branes and matrices}

\author{Ian T. Ellwood\\{Department of Physics,
             University of Wisconsin,\\ 
             Madison, WI 53706, U.S.A.}\\{iellwood@physics.wisc.edu}}

\abstract{
We construct a general map between a Dp-brane with magnetic flux and a
matrix configuration of D0-branes, by showing how one can
rewrite the boundary state of the Dp-brane in terms of its D0-brane
constituents.  This map gives a simple prescription for constructing
the matrices of fuzzy spaces corresponding to branes of arbitrary
shape and topology.  Since we explicitly identify the D0-brane degrees
of freedom on the brane, we also derive the D0-brane charge of the
brane in a very direct way including the $\widehat{A}$-genus term. As
a check on our formalism, we use our map to derive the
abelian-Born-Infeld equations of motion from the action of the
D0-brane matrices.
}


\preprint{MAD-TH-04-13, hep-th/0501086}

\begin{document}

\section{Introduction}

It has been known since the 80's that it is possible to encode the
geometry of surfaces embedded in flat space using a collection of
matrices \cite{MatrixReg}.  While the original motivation of
introducing such a matrix description was to regularize the membrane
action, it was found in
\cite{Banks:1996vh,Taylor:2001vb,Taylor:1999gq,Taylor:1999pr,Myers:1999ps} that these fuzzy
branes appear naturally in string theory  as excitations of D0-branes.

In spite of the impressive number of examples of fuzzy geometries
formed out of collections of D0-branes
\cite{Kabat:1997im,Castelino:1997rv,Taylor:1997dy,Ishibashi:1998ni,Myers:2003bw,Bak:2001kq,
Maldacena:2002rb,Hashimoto:2003pu,Hashimoto:2004fa,Karabali:2004xq},
it has never been completely clear what the precise map is between a
given Dp-brane configuration and its associated matrix description
except in cases with a high degree of symmetry.

In principle, such a map can be derived from the Seiberg-Witten map
\cite{Seiberg:1999vs}.  This viewpoint has been emphasized in a series
of papers by Cornalba, Schiappa and Bordalo~\cite{Cornalba:2000wq}.
In this approach, one uses the Seiberg-Witten map to replace the
algebra of functions on the brane with a star algebra.  This star
algebra in turn can be replaced by an operator algebra, which can then
be interpreted as the algebra of the matrix positions of the
D0-branes.  The drawback of this method is that defining a star
product on a curved brane can be quite complex~\cite{Kontsevich}.

What we will show in this paper is that there is actually a more
direct method for constructing the D0-brane matrices which correspond
to a given Dp-brane background.  To derive this map,
we take the boundary state of a Dp-brane and manipulate it into
the form of the boundary state of a collection of D0-branes.  
For the case of a flat Dp-brane with uniform flux, an equivalent
calculation was performed in \cite{Ishibashi:1998ni}, where the precise
equivalence of the Dp-brane and D0-brane descriptions was
demonstrated.  

When the Dp-brane is curved, the discussion is somewhat
more subtle.
To see why, consider the case of a compact Dp-brane with $N$ units of
D0-brane charge.  In the dual D0-brane description, the positions of the
D0-branes are given by $N\times N$ matrices that form a fuzzy
version of the original Dp-brane.  However, since the matrices are of
finite size, there is a mismatch between the infinite number of Dp-brane
configurations and the finite number of matrix configurations.  This mismatch
is most severe when $N =1$.  In this case, the matrices degenerate to
scalars representing the location of the single D0-brane and it is 
not even possible to form a fuzzy geometry.
This discrepancy between the Dp-brane picture and the D0-brane picture
implies that the exact duality derived in the flat case cannot hold
in the general curved case without some caveats .

To understand the origin of this issue, we need to study the properties of string
endpoints in the presence of a magnetic field.  As has been discussed
in \cite{Bigatti:1999iz,Brodie:2000yz,Fabinger:2002bk}, the
noncommutative geometry of the D0-branes arises in the
Dp-brane picture from the coupling of the string endpoints to
the brane geometry.  Indeed, the endpoint action is just the ordinary
coupling of a charged particle to a background magnetic field.
Quantizing this endpoint action gives the usual Landau levels.  
If  the magnetic field is strong, the separation between the
levels will be large, and we can assume that the endpoint is constrained to the
lowest Landau level.  As is well known, constraining a theory to the lowest 
Landau level naturally leads to  a noncommutative geometry 
\cite{Karabali:2004xq}, which, in this case, we can
interpret as the fuzzy geometry of the D0-branes.

Unfortunately, this argument does not explain why the duality shown in
\cite{Ishibashi:1998ni} works even when the field strength is small.
Here the story is more subtle and we give most of the details in
section \ref{s:decoupling}.  Consider quantizing the world sheet
action of a string ending on a flat brane, with uniform magnetic flux.  To study the
behavior of the endpoint in an unambiguous way, we must impose a cutoff.  
In section \ref{s:decoupling}, we will
consider a lattice regularization in which the endpoint becomes a
single lattice site.

It is easy to check in such a model that the endpoint has a non-vanishing
expectation value to leave the lowest Landau level. However,
all of the physics associated with the excited states of the endpoint lives
at the cutoff scale.  When the cutoff is removed these excited states
decouple from the theory.  Hence, even when the background
field strength is small, we may assume the endpoint to be constrained to
its ground state.

We can now ask whether this same mechanism will work on a curved
brane.  For a curved brane, we must worry about the potential
divergences which arise from the coupling of the endpoint to the
curved surface of the brane.  These divergences, which also live at
the cutoff scale, must be removed by renormalization.  In general, the
decoupling of the endpoint excitations described above will be
destroyed by this renormalization procedure.  This will allow the
string to see higher Landau levels of the endpoint and, hence, a more
complete picture of the brane.  Non-trivial Dp-branes with one unit of
D0-brane charge must be of this form.  We will refer to such branes as
smooth branes to distinguish them from fuzzy branes.

On the other hand, for special regulators, the divergences from the
curvature of the brane will not interfere with the decoupling of the
endpoint excitations.  For these branes, the endpoint may be taken to
live in the lowest Landau level and much of the original geometry of
the brane will be lost.  These branes are naturally identified with
fuzzy configurations of D0-branes.
As we will see, when we perform our boundary state computation, such a
regulation can be found.  However, we will also argue, that, for a
different choice of regulator, the higher modes of the endpoint would
not decouple.

We can now make very precise what we mean by relating a brane
configuration and matrix configuration.  A brane/matrix map simply
relates the world sheet theory on a Dp-brane in which the endpoint
excitations have decoupled, to a world sheet theory on a matrix
configuration of D0-branes.  In this sense, the map works for any
field strength.  Of course if we wish to make sense of the smooth
branes, we can only apply the map in cases where the field strength is
large or in special cases like the flat Dp-brane with uniform flux.

We now describe the map explicitly.  In light of the well known relationship
between fuzzy D0-brane configurations and the projection onto
 to the lowest Landau level, the form of the
map should not be surprising.  However, projecting an observable onto
the lowest Landau level is an ambiguous procedure, and the boundary state
computation gives a preferred choice.  A different method of projecting
onto the lowest Landau level is discussed, for example, in \cite{Karabali:2004xq}.
Our map can be described as follows:
\begin{enumerate}
\item Let the world volume Dirac operator in the presence of $A_\mu$
be denoted $\mathcal{D}$, and its zero modes given by $\mathcal{D}
|\rho\rangle =0$.  Furthermore, pick a basis in which the zero modes
are orthonormal.
\item Let the coordinates on the Dp-brane be denoted $\xi^\alpha$ and
the embedding into the target space be given by $x^M(\xi)$.  Then the
coordinate matrices of the D0-branes are given by
$\mathbf{X}^M_{\rho_1 \rho_2} = \langle \rho_1 | x^M (\xi) |\rho_2
\rangle$.
\end{enumerate}
Here $M$ is assumed to be a spacial index.  Generalizing this
construction to time dependent configurations and configurations in
which the dual description is a spacially dependent matrix is not
hard and will be discussed in section \ref{s:timedependent}.

This map is, as expected, nothing but a projection of the position
operators onto the lowest Landau level.  The fact that the lowest
energy states are zero modes of the Dirac operator is just a
consequence of world sheet supersymmetry.  The one subtlety in this
map is that some of the zero modes correspond not to D0-branes, but to
D$\bar{0}$-branes.  As we will show later, the D0-branes correspond to
positive chirality modes, whereas, the D$\bar{0}$-branes correspond to
negative chirality modes.  Note that this correspondence gives an easy
derivation of the D0-brane charge formula; the D0-brane charge is
simply the index of the Dirac operator.  This reproduces the usual
formula found in
\cite{Green:1996dd,Cheung:1997az,Minasian:1997mm,Freed:1999vc}, in the
case when the background spacetime is flat. For other methods of
reproducing the D0-brane charge from a boundary state computation,
see \cite{DiVecchia:1999uf,Asakawa:2003ax}.

After deriving this correspondence, we perform a simple check on our
results.  Starting with the action for a matrix configuration of
D0-branes, we use our map to reproduce the usual abelian-Born-Infeld
equations of motion.  This requires explicitly computing the map in
the case of a Dp-brane that has small curvature and a slowly varying
field strength.  The resulting expressions reproduce the usual
formulas for this case
\cite{Ishibashi:1998ni,Seiberg:1999vs,Cornalba:2000wq,miao1996,Hyakutake:2001kn}.

The plan of the paper is as follows: In section
\ref{s:boundarystatebosonic}, we begin by deriving the brane/matrix
map for the simpler case of the bosonic string. In section
\ref{s:boundarystateSUSY}, we extend the computation to the
superstring.  In section \ref{s:BI}, we use our map to compute the
abelian-Born-Infeld equations of motion from the action of the
D0-brane matrices.  In section \ref{s:decoupling}, we discuss the
decoupling of of the higher modes of the string endpoint.  We
conclude, in section \ref{s:conclusions}, with some possible future
directions.

\section{Boundary state analysis}\label{s:boundarystatebosonic}

In this section, we derive our brane/matrix map using a manipulation
of the boundary state of a Dp-brane.  For simplicity, we start with
the simpler bosonic argument.  We generalize to the superstring in the
next section.  Many of the techniques used are similar to those in
\cite{Ishibashi:1998ni,Kraus:2000nj,Asakawa:2003ax}.

\subsection{Bosonic boundary state basics}

We begin with a few basic facts about bosonic boundary states for flat
branes.  If we imagine a string being emitted from a brane, the
boundary of the string world-sheet is simply a loop, $x^M(\sigma)$,
living on the brane's surface, where $\sigma$ runs from $0$ to $2\pi$.
A basic prescription for computing the boundary state
\cite{Callan:1987px} is to sum over the string states representing
every such loop, giving each state a weight $e^{-S_B}$, where $S_B$ is
the boundary action of the loop.

Consider a flat brane with Neumann directions, $i$, and Dirichlet
directions, $a$.  Let $X^M(\sigma)$ be the world sheet field,
$X^M(z,\bar{z})$, evaluated at $e^{i \sigma}$. The matter part of the
Dp-brane boundary state is given, up to normalization, by
\cite{Callan:1987px}
\begin{equation}\label{eq:bosonicdp}
  |\mathcal{B}\rangle = \int \mathcal{D}x^i(\sigma) e^{-S_B(x)}\,
   |x^i(\sigma); X^a(\sigma) = 0 \rangle,
\end{equation}
where the state, $|x^i(\sigma);X^a= 0\rangle$, is an eigenvector of 
$X^M(\sigma)$ with eigenvalues
\begin{align}
   X^i(\sigma) |x^i(\sigma); X^a = 0 \rangle
    &= x^i(\sigma) |x^i(\sigma); X^a = 0 \rangle \label{eq:prop1},\\
  X^a(\sigma) |x^i(\sigma); X^a = 0 \rangle \label{eq:prop2}
    &= 0,
\end{align} 
and the boundary action is given by
\[
  S_B(x) = \int d\sigma\, A_i(x) \partial_\sigma x^i.
\]

We now suppose that our Dp-brane is given by some submanifold, $Q$,
embedded in flat space.  Let the coordinates on $Q$ be given by
$\xi^\alpha$ and let $x^M(\xi^\alpha)$ be the embedding of $Q$ into
the background spacetime.  We take the $M$ index to run only over
spacial indices and the configuration is assumed to be static. We will
generalize to dynamical branes later.  As with the flat case, there
are standard formulas 
\cite{Asakawa:2003ax,Kraus:2000nj,Callan:1987px,Hashimoto:1999dq} 
for the boundary state
of such a brane. We use a slightly different notation, however, which
aids our analysis.  We have
\begin{equation}\label{eq:bs2}
  |\mathcal{B}\rangle = \int \mathcal{D}\xi(\sigma) \, e^{-S_B(\xi)}
  |x^M (\xi(\sigma))\rangle,
\end{equation}
where the state, $|x^M (\xi(\sigma))\rangle$, satisfies
\begin{equation}
  X^M(\sigma) |x^M (\xi(\sigma))\rangle
 = x^M(\xi(\sigma)) |x^M (\xi(\sigma))\rangle,
\end{equation}
and the boundary action is now given by
\[
  S_B = \int A_\alpha(\xi) \partial_\sigma \xi^\alpha.
\]
To proceed, rewrite the state, $|x^M (\xi(\sigma))\rangle$, in
momentum space
\begin{equation}\label{eq:postomom}
  |x^M (\xi(\sigma))\rangle = \int \mathcal{D}
 \pi_M(\sigma)\, \exp\left( i \int d\sigma \, x^M(\xi) \pi_M
 \right) |\pi_M(\sigma)\rangle,
\end{equation}
where $|\pi_M(\sigma)\rangle$ is an eigenvector of the conjugate momenta, 
$\Pi^M(\sigma) = -i \delta/\delta X^M(\sigma)$,
with eigenvalues
\[
  \Pi_M(\sigma) |\pi_M(\sigma)\rangle
 = \pi_M(\sigma) |\pi_M(\sigma)\rangle.
\]
Substituting equation (\ref{eq:postomom}) into equation
(\ref{eq:bs2}), we can rewrite the boundary state as
\begin{equation} \label{eq:lagrangeform}
  |\mathcal{B}\rangle
 = \int \mathcal{D} \xi(\sigma) \mathcal{D} \pi(\sigma)
   e^{-S_B+ i \int d\sigma \, x^M(\xi) \pi_M(\sigma)}
  |\pi_M(\sigma)\rangle.
\end{equation}
The next step in our analysis is to rewrite the path integral over the
field, $\xi^\alpha$, in a more illuminating way.  Unfortunately, as
written, this path integral is not well defined as there is no kinetic
term for $\xi^\alpha$.  This problem is a result of our switch from an
$X^M$ eigenbasis to a $\Pi_M$ eigenbasis.  As a simple example of this
phenomenon, consider the integral
\[
  \int_{-\infty}^\infty d\pi dx\, e^{ix\pi} e^{-\pi^2/2}.
\]
Attempting to perform the $x$ integral first is ambiguous.  However,
performing the $\pi$ integral gives
\[
   \int_{-\infty}^\infty \,dx\, \sqrt{2\pi}\, e^{-x^2/2},
\]
which is perfectly well defined.  

Similarly, in our boundary state computation, if we performed the
$\pi_M(\sigma)$ integral first, we would just get back to the $X^M$
basis $|x^M(\sigma)\rangle$.  Examining the explicit oscillator
representation of $|x^M(\sigma)\rangle$, given in
\cite{Callan:1987px}, one finds a leading term $\sim e^{-x^2}$, which
makes the $\xi^\alpha$ integral well behaved.

To get around this problem, we regularize the $\xi^\alpha$ path integral
 by adding a small kinetic term to our boundary action
\begin{equation}
   S_B = \int d\sigma \bigl( \frac{1}{2}
  \left( \epsilon\right) 
\partial_\sigma \xi^\alpha \partial_\sigma
\xi^\beta g_{\alpha\beta} 
+ A_\alpha(\xi) \partial_\sigma \xi^\alpha
\bigr).
\end{equation}
This simply gives a small mass, $\epsilon$, to our string endpoint
which we will limit to zero at the end of the computation.  Note that
$\epsilon$ need not depend on the regulator we use to make boundary
state finite.  For example equation (\ref{eq:lagrangeform}) can me
made finite by imposing a cutoff on the $\pi$ path integral.  From
this point of view, we have merely added something to our integral to
make it better defined and we will limit it away at the end of the
calculation.  Equivalently, one can think of this small kinetic term
as defining the measure for the $\xi$ integral in a natural, covariant
way.

On the other hand, there is a very real sense in which $\epsilon$
could be thought of as depending on the cutoff. It is easy to
construct a short distance cutoff in which the endpoint of the string
obtains a mass proportional to the cutoff.  In such a scenario, the
limit as $\epsilon$ and the cutoff go to zero will
produce many new terms in the boundary state which depend on excited
states of the endpoint.

We will not consider this possibility here, although it is interesting,
and, if done properly, might give some insight into the boundary state of
smooth branes.  For now, however, we will consider $\epsilon$ a parameter
which is limited to zero at fixed cutoff.

\subsection{Interpreting in terms of D0-branes}

To express the boundary state in terms of D0-brane degrees of freedom,
we rewrite the path integral over $\xi^\alpha$ in terms of a trace over the
Hamiltonian.  Putting
\begin{equation}
  \mathcal{L} =  \frac{1}{2} \left(\epsilon\right)
    \partial_\sigma \xi^\alpha \partial_\sigma \xi^\beta g_{\alpha\beta}
   +
    A_\alpha(\xi) \partial_\sigma \xi^\alpha
   +  
    i x^M(\xi) \pi_M(\sigma),
\end{equation}
the Hamiltonian is given by
\begin{equation}
  \mathcal{H} = \frac{1}{2\epsilon} 
(P^\alpha-A^\alpha)g_{\alpha \beta}(P^\beta-A^\beta)
- i x^M(\xi) \pi_M(\sigma).
\end{equation}
We can then use the standard identity,
\begin{equation} \label{eq:ltoh}
  \int \mathcal{D}\, \xi e^{-S}
 = \text{TrP} e^{-\int{d\sigma \mathcal{H}(\sigma)}},
\end{equation}
where $\text{P}$ denotes path ordering and the trace, $\text{Tr}$, is over the
the space of states.  It is convenient to define
\[
  \mathcal{H}_0 = \frac{1}{2}
(P^\alpha-A^\alpha)g_{\alpha \beta}(P^\beta-A^\beta),
\]
so that $\mathcal{H} = \mathcal{H}_0/\epsilon - i x^M (\xi) \pi_M$.
Also, it is useful to pick a basis in which $\mathcal{H}_0$ is diagonal,
\[
    \mathcal{H}_0 |\lambda\rangle = E_\lambda|\lambda\rangle.
\]

Note that, at the quantum level, there is an ordering ambiguity in
how one defines the Hamiltonian.  For the superstring, this will
be less of a problem, as the Hamiltonian is naturally written
as the square of the Dirac operator.  Even in the bosonic case,
though, covariance puts some constraints on the operator
form of $\mathcal{H}_0$.

Using equation (\ref{eq:ltoh}), we can rewrite our path integral as
\[
   \text{TrP} \exp
   \left\{ 
      -\int d\sigma  (\mathcal{H}_0/ \epsilon - i x^M \pi_M )  
   \right\}.
\]
Consider expanding out the exponent in powers of $x^M \pi_M$.  For
illustration purposes, we do this for the second order term.  The
manipulations for the other terms are similar.  Taking only two powers
of $x^M \pi_M$, we get the expression
\begin{multline} \label{eq:secondorderterm}
  \sum_\lambda \int_0^{2\pi} d\sigma_1 d\sigma_2\,\, \theta(\sigma_2 -
\sigma_1)  \langle \lambda | e^{-(2\pi -\sigma_2)
\mathcal{H}_0/\epsilon} \left(i x^{M_2} \pi_{M_2}(\sigma_2)\right)
e^{-(\sigma_2-\sigma_1) \mathcal{H}_0/\epsilon} \\ \left(i x^{M_1}
\pi_{M_1}(\sigma_1)\right) e^{-\sigma_1\mathcal{H}_0/\epsilon}
|\lambda \rangle,
\end{multline}
where $\theta(\sigma_2-\sigma_1)$ is a step function which enforces the
path ordering.
Consider the effect of the various insertions of 
\[
   e^{-\delta \sigma \mathcal{H}_0/ \epsilon}
    = \sum_{\lambda} 
    |\lambda\rangle 
       e^{-\delta \sigma E_\lambda/ \epsilon } \langle \lambda |.
\]
Note that as $\epsilon \to 0 $, the term $e^{-\delta \sigma E_\lambda/
\epsilon } $ vanishes unless $E_\lambda = 0$.  Hence in the small $\epsilon$
limit, we can write
\begin{equation} \label{eq:Hasprojector}
    \lim_{\epsilon \to 0}e^{-\delta \sigma \mathcal{H}_0/ \epsilon}
     = \sum_{\rho } |\rho\rangle \langle \rho|,
\end{equation}
where the $|\rho\rangle$ are the states with zero
energy. Unfortunately, in most interesting cases there are, in fact, no
states of zero energy.  However, one can fix this problem by writing
\[
   \mathcal{H}_0 = \mathcal{H}_0' + \mathcal{E}_0,
\]
where $\mathcal{E}_0$ is the energy of lowest state and $\mathcal{H}'$
now has an eigenstate with zero energy.  One can then perform the
calculation as before,  using $\mathcal{H}'$ instead of
$\mathcal{H}$, since the constant term only affects the overall
normalization of the boundary state.   
As we are not dealing with the issue of finding the correct
normalization of the boundary state, we will ignore this point and
simply assume at least one state with zero energy.  When we work with
the superstring, we will see that there are {\em typically} states
with zero energy.  

Another potential problem arises if there is a continuum of energies
instead of a gap above the ground state energy.  In this case, the
limit $\epsilon \to 0$ is not well defined and our analysis is not
legitimate.  This occurs, for example, in the case of an infinite flat
D2-brane with no flux, where there is no D0-brane interpretation.

Using the formula (\ref{eq:Hasprojector}) in our expression
(\ref{eq:secondorderterm}), we get
\begin{equation} \label{eq:secondorderterm2}
  \sum_{\rho_1 \rho_2} \int_0^{2\pi} 
d\sigma_1 d\sigma_2 \theta(\sigma_2 - \sigma_1) 
  \langle \rho_1 |
 i x^{M_2} \pi_{M_2}(\sigma_2)|\rho_2\rangle
 \langle \rho_2 |i x^{M_1} \pi_{M_1}(\sigma_1)
  |\rho_1 \rangle,
\end{equation}
where, as before, the $|\rho_i\rangle$ are assumed to have zero energy.
To put this expression in a simpler form, define the matrices
\begin{equation}\label{eq:matrixdef}
   \mathbf{X}^M_{\rho_1\rho_2} = 
   \langle \rho_1 | x^M |\rho_2\rangle.
\end{equation}
Equation (\ref{eq:secondorderterm2}) can then be written as
\begin{equation}
(i)^2\int_0^{2\pi} d\sigma_1 d\sigma_2 \,\,\theta(\sigma_2 - \sigma_1) 
  \text{tr}\left[\mathbf{X}^{M_2} \mathbf{X}^{M_1} \right]
  \pi_{M_2}(\sigma_2) \pi_{M_1}(\sigma_1)
  |\lambda_1 \rangle,
\end{equation}
which is just the second order term in
\[
  \text{trP} \exp \left\{ i \int d \sigma\, \mathbf{X}^M \pi_M(\sigma) \right\}.
\]
The other terms in the expansion follow from a similar argument.  The
full boundary state is given by
\begin{equation}
  \int \mathcal{D} \pi(\sigma) \,\,
  \text{trP} \exp \left\{ i \int d \sigma\, 
\mathbf{X}^M \pi_M(\sigma) \right\} 
  |\pi^M\rangle.
\end{equation}
Switching back to the $X^M$ basis gives our final expression
\[
  | \mathcal{B}\rangle = 
    \text{trP} \exp \left\{ i \int d \sigma\, 
\mathbf{X}^M \Pi_M(\sigma) \right\}
    |x^M = 0\rangle.
\]
This expression is the usual formula for the boundary state of a
collection of bosonic D0-branes with matrix-valued positions
$\mathbf{X}^M$ 
\cite{Asakawa:2003ax,Kraus:2000nj,Callan:1987px,Hashimoto:1999dq}.  
Note that we now have an
explicit map between the brane configuration and the
matrix-configuration of the D0-branes given by
eq. (\ref{eq:matrixdef}).  We now present two familiar examples to
illustrate the map.

\subsubsection*{Example 1: Constant $F_{ij}$ on a flat D2-brane}

Consider, as an example, a flat D2-brane with constant $F_{12} = f$.
For this case, the equivalence of the D2-brane boundary state and the
D0-brane boundary state was shown in \cite{Ishibashi:1998ni}.  We can
take the gauge field to be $A_1 = -f X^2$ with $A_2 = 0$.  Note that
the Hamiltonian, $\mathcal{H}_0$ is simply the Hamiltonian for a
spinless particle in a constant magnetic field.  This is just the
ordinary Landau problem.  The states in the lowest energy level are
labeled by $|k\rangle$.  The matrices $\mathbf{X}^1$ and
$\mathbf{X}^2$ are given by
\begin{align}
  \mathbf{X}^1_{kk'} &= -i\frac{\partial}{\partial k} \delta(k-k'),\\
  \mathbf{X}^2_{kk'} &= \frac{k}{F} \delta(k-k').
\end{align}
Hence $[\mathbf{X}^2,\mathbf{X}^1] = \frac{i}{f}\, \mathbf{1} $, where
$\mathbf{1}$ is the identity matrix.  This is the standard
configuration of D0-branes said to correspond to a D2-brane.

\subsubsection*{Example 2: Spherical D2-brane with uniform flux}

Now consider a spherical D2-brane with constant flux.  For this setup,
$\mathcal{H}_0$ is given by the Hamiltonian for a spinless particle
constrained to a sphere with a constant magnetic field pointing
radially out from the sphere.  This problem is quite old, and was
solved, for example, in \cite{Fierz}.  A more elegant solution is
presented in \cite{Haldane:1983xm} and a modern discussion is given in
\cite{Fabinger:2002bk}.  For our purposes we need only know that the
Hamiltonian takes the form
\[
\mathcal{H}_0 = \frac{2\pi\alpha'}{2 R^2} \left(L^2 -\frac{N^2}{4}\right),
\]
where the $L^i$ are the generators of angular momentum in the presence
of the gauge field and $i = 1,2,3$.  The ground state is an $N+1$
dimensional representation of the $L$s and we can compute the matrices
$\mathbf{X}^i$ using the Wigner Eckart theorem.  Let the ground states
be denoted $|\ell = N/2,m\rangle$, with $-N/2\le m\le N/2$.  We have
\begin{equation}
 \mathbf{X}^i_{m_1 m_2} = \langle \ell, m_1| x^i |\ell,m_2 \rangle
 = \frac{\langle \ell,m_1| \sum_i L^i x^i |\ell,m_1\rangle}{\ell(\ell+1)}
\langle \ell m_1| L^i |\ell m_2\rangle.
\end{equation}
Note the matrix element $\langle \ell,m_1| \sum_i L^i x^i |\ell,m_1\rangle$
is independent of $m_1$. An explicit computation gives
\[
  \langle \ell,m_1| \sum_i L^i x^i |\ell,m_1\rangle
 = R\, \frac{ N}{2}.
\]
Since $\langle \ell m_1| L^i |\ell m_2\rangle$ is just the matrix
representation of $L^i$, we can write
\[
  \mathbf{X}^i =  \frac{R}{(N/2+1)}\, \mathbf{J}^i,
\]
where the $\mathbf{J}^i$ are the generators of the $N+1$ dimensional
representation of $SU(2)$.
This is the standard representation of a sphere using D0-branes
\cite{Taylor:2001vb,Myers:2003bw}.

Note that the dimension of the representation is $N+1$, implying that
a Dp-brane with $N$ units of flux is built from $N+1$ D0-branes in the
bosonic theory.  This disagrees with the expected result from the
superstring, but, since there is no charge associated with the
D0-brane in the bosonic theory, it is not inconsistent.  We will see
that, in the superstring analysis, we always reproduce the correct
D0-brane charge.

\subsection{Time dependent backgrounds}\label{s:timedependent}

The only new ingredient in a time dependent background is that the 
Hamiltonian, $\mathcal{H}_0$, now depends on $X^0$.  This implies that
we must make the replacement
\[
  e^{-\delta \sigma \mathcal{H}_0/\epsilon} \to P \exp
  \left\{-\int_{\sigma_1}^{\sigma_2} d\sigma\,\,
  \mathcal{H}_0(X^0(\sigma))/\epsilon\right\}.
\]
For a general time dependent Hamiltonian, we cannot evaluate this
object.  However, if we assume that we can follow the eigenvectors of the
Hamiltonian in time and that no eigenvectors appear or disappear,
we can rewrite this expression as follows:  Label the eigenvectors
in the ground state at time $X^0$ by $|\rho(X^0) \rangle$.  Then using
the adiabatic theorem we get
\begin{equation}
  P \exp
  \left\{-\int_{\sigma_1}^{\sigma_2} d\sigma\,\,
  \mathcal{H}_0(X^0(\sigma))/\epsilon\right\}
= \sum_\rho |\rho(X^0(\sigma_2))\rangle \langle \rho(X^0(\sigma_1))|.
\end{equation}
The difference between this formula and the time independent formula (\ref{eq:Hasprojector}),
can be accounted for by modifying the definition of the matrices, 
\[
  \mathbf{X}^M_{\rho_2 \rho_1}(X^0) = 
   \langle \rho_2(X^0)|x^M| \rho_1(X^0)\rangle,
\]
so that the matrices now depend on time.  We could also consider cases
where $\mathcal{H}_0$ depends on some spacial direction.  The same
argument as above will then work, giving spacially dependent matrices.

\section{Superstring boundary states}\label{s:boundarystateSUSY}

In this section, we generalize the results of the previous section to
the superstring.  We begin, as before, with a review of the form of the
boundary state and then show how to rewrite it in terms of D0-branes.
Much of the discussion is parallel with the bosonic case, however, we
find a new contribution when two operators collide in the path
integral.

\subsection{The SUSY boundary state for flat branes}

To include the fermion fields, we define boundary fields, $\Psi^M
(\sigma)$ and $\widetilde{\Psi}^M (\sigma)$, to be the fermion fields,
$\Psi(z)$ and $\widetilde{\Psi}(\bar{z})$, at $z = e^{i\sigma}$ and
$\bar{z} = e^{-i\sigma}$ respectively.  We then define the field,
$\Psi^M_{\pm} = \Psi^M\pm i \widetilde{\Psi}^M$.  Before
defining the complete boundary state, it is standard to define
\cite{Callan:1987px}
\[
  |\mathcal{B};\pm\rangle = 
   \int \mathcal{D}x^i \mathcal{D} \psi^i 
       e^{-S_B(x,\psi)}|x^i;X^a = 0\rangle 
       |\psi^i;\psi^a = 0, \pm\rangle,
\]
where the state $|\psi^i;\psi^a = 0;\pm\rangle$, which is a function of 
the classical Grassmann field $\psi^i(\sigma)$, has
the properties
\begin{align}
     \Psi_\pm^i(\sigma)|\psi^i;\psi^a = 0;\pm\rangle
      &= \psi^i(\sigma)|\psi^i;\psi^a = 0;\pm\rangle, 
\\
     \Psi_\pm^a(\sigma)|\psi^i;\psi^a = 0;\pm\rangle
      &= 0.
\end{align}
The boundary action, $S_B$, is given by
\[
  S_B = \int d\sigma \, \left(A_i\partial_\sigma x^i
-i\frac{1}{2} F_{ij} \psi^i\psi^j\right).
\]
The complete boundary state is then given by \cite{Callan:1987px}
\[
  |\mathcal{B}\rangle
 = P\widetilde{P}_+ |\mathcal{B};+\rangle_{\text{NS}}
  +P\widetilde{P}_- |\mathcal{B};+\rangle_{\text{RR}},
\]
where $P$ and $\tilde{P}$ are GSO projectors.  Since we will not be concerned with
the GSO projections, we will simply work with $|\mathcal{B};+\rangle$.
For convenience, we will also drop the $\pm$ labels as they will
not play any roll in the discussion.

\subsection{The boundary state for curved branes}
Generalizing to curved branes is straightforward.  For a nice
discussion, see \cite{Asakawa:2003ax,Kraus:2000nj,Hashimoto:1999dq}.  As
before we take the coordinates on the world volume of the brane to be
$\xi^\alpha$ and we take the spinors, $\psi^i$, to live on the world
volume, with $i$ a tangent space index.  We take, as before,
$x^M(\xi)$ to be the embedding of the brane into spacetime, and
$g_{\alpha\beta}(\xi)$ to be the induced metric on the brane.
Finally, we introduce a vielbein ${e^\alpha}_i(\xi)$.  We can then
define the boundary state as
\[
  \int \mathcal{D}\xi^\alpha\,\mathcal{D}\psi^\alpha\,
  e^{-S_B}
  |x^M(\xi)\rangle |\Psi^M = 
 \frac{\partial x^M}{\partial \xi^\alpha}e^\alpha_i(\xi) \psi^i\rangle,
\]
where
\[
  S_B = \int d\sigma 
  \left( 
     A_\alpha(\xi) \partial_\sigma \xi^\alpha
     +i\frac{1}{2} F_{\alpha\beta} {e^\alpha}_i {e^\alpha}_j \psi^i \psi^j
  \right).
\]
It is useful to rewrite the state, $|\Psi^M = \frac{\partial
x^M}{\partial \xi^\alpha}e^\alpha_i(\xi) \psi^i\rangle$, in
Fourier space:
\begin{equation}
  |\Psi^M = \frac{\partial x^M}{\partial \xi^\alpha}
    e^\alpha_i(\xi) \psi^i\rangle
=
  \int \mathcal{D} \chi^M 
  e^{- \int d\sigma \, \frac{\partial x^M}{\partial \xi^\alpha}
   e^\alpha_i(\xi) \psi^i
\chi_M} |\chi_M\rangle,
\end{equation}
where the state, $|\chi_M\rangle$, satisfies
\[
                       \Psi^M(\sigma) |\chi_M\rangle 
= \frac{\delta}{\delta \chi^M(\sigma)}|\chi_M\rangle.
\]
This allows us to write
\begin{equation}
  |\mathcal{B}\rangle
 = \int \mathcal{D}\xi^\alpha\,\mathcal{D}\psi^\alpha\,
\mathcal{D} \chi^M\, \mathcal{D} \pi^M \, 
e^{
       -S_B+ \int d\sigma\,
   \left(
      i x^M \pi_M  - 
   \psi^i {e^\alpha}_i \frac{\partial x^M}{\partial \xi^\alpha}
    \chi_M
   \right)
}
|\pi^M\rangle |\chi^M\rangle.
\end{equation}

We would now like to perform the path integral over the $\xi$ and
$\psi$ fields, but, as in the bosonic case, the path integral is not
well-defined.  As before, we fix this problem by adding a
small kinetic term for $\xi^\alpha$ and $\psi^i$.  Our exponent
becomes $-\int d\sigma \mathcal{L}$, where $\mathcal{L}$ is given by
\begin{multline}\label{eq:lagrangian}
\mathcal{L} = 
\frac{1}{2} \left(\epsilon\right)
\partial_\sigma \xi^\alpha \partial_\sigma \xi\beta g_{\alpha\beta}
+
i \frac{1}{2} \left(\epsilon\right)
\psi^i (\delta_{ij}\partial_\sigma - \partial_\sigma \xi^\alpha
 \omega_{\alpha ij} 
)\psi^j 
\\+
A_\alpha \partial_\sigma \xi^\alpha
              + i \frac{1}{2} F_{\alpha\beta} \psi^\alpha\psi^\beta
     -i x^M \pi_M + \psi^i {e^{\alpha}}_i 
     \frac{\partial x^M}{\partial \xi^\alpha}\chi_M.
\end{multline}
Note that we have introduced the spin connection,
$\omega_{\alpha}^{ij}$, in order to make a covariant kinetic term for
$\psi^i$.  This action is simply the supersymmetric extension of the
original action.  We now rewrite the path integral over $\xi$ and
$\psi$ in the Hamiltonian formalism as we did in the bosonic case.

\subsection{Interpreting the boundary state in terms of D0-branes}

As we did in the bosonic case, we rewrite our path integral as
\begin{equation} \label{eq:expofh}
\text{PTr} e^{-\int d\sigma \mathcal{H}},
\end{equation}
where the Hamiltonian, $\mathcal{H}$, corresponding to the Lagrangian (\ref{eq:lagrangian}) is
given by
\[
\mathcal{H} = 
  \frac{1}{2 \epsilon} \mathcal{D}^2 -
 i x^M \pi_M + \frac{1}{\sqrt{2\epsilon}}\Gamma^i {e^{\alpha}}_i 
     \frac{\partial x^M}{\partial \xi^\alpha}\chi_M.
\]
Here $\mathcal{D}$ is the ordinary Dirac operator,
\[
 \mathcal{D} = \Gamma^I(-i\partial_I - A_I),
\]
 and the $\Gamma^I$
satisfy $\{\Gamma^I,\Gamma^J\} = 2 \delta^{IJ}$ as usual.  The factor
of $1/\sqrt{2\epsilon}$ comes from our non-standard normalization of
the fermion kinetic term.  As in the bosonic case, define
\[
  \mathcal{H}_0 = \frac{1}{2} \mathcal{D}^2.
\]
Let the states, $|\lambda\rangle$, be the eigenstates of $\mathcal{H}_0$,
so that $\mathcal{H}_0 |\lambda\rangle = E_\lambda |\lambda\rangle$.  As
we found in the bosonic case, the Hamiltonian will project onto the
states of zero energy.  As before, denote the states of zero energy by
$|\rho\rangle$.  Since $\mathcal{H}_0$ is just the square of the Dirac
operator, we have
\[
   \mathcal{D} |\rho\rangle = 0.
\]
For the term, $i x^M \pi_M$, the analysis proceeds in the same way as
for the bosonic case.  The $1/\sqrt{2\epsilon}$ term,
however, needs to be treated in a slightly different way.  It is
convenient to note that
\[
  \,\Gamma^i {e^{\alpha}}_i 
     \frac{\partial x^M}{\partial \xi^\alpha}\chi_M
 = \,[\mathcal{D},x^M (\xi)] \chi_M,
\]
which implies that this term vanishes between states of zero energy,
\[
  \langle \rho| \frac{1}{\sqrt{2\epsilon}}\,[\mathcal{D},x^M (\xi)] \chi_M
 |\rho' \rangle = 0,
\]
as the Dirac operator kills either the state on the left or on the
right.  Since, as in the bosonic case, one will have a factors of
$e^{-\delta \sigma \mathcal{H}_0/\epsilon}$ around each insertion of $
1/\sqrt{2\epsilon}\,[\mathcal{D},x^M (\xi)] \chi_M$, and
these factors project onto the ground state, one might suppose that this
term will not contribute. Indeed, when the term is separated from
other such insertions, this is true. However, we get a nonzero contribution when two
$1/\sqrt{2\epsilon}\,[\mathcal{D},x^M (\xi)] \chi_M$ terms
collide.  

Consider a term in the expansion of the exponential (\ref{eq:expofh}), with two
neighboring insertions of $
1/\sqrt{2\epsilon}\,[\mathcal{D},x^M (\xi)] \chi_M$, which
are separated by $\Delta \sigma = \sigma_2-\sigma_1$
\begin{equation}
1/\sqrt{2\epsilon}\,[\mathcal{D},x^{M_1} ] \chi_{M_1}
  (\sigma_2)
 \left(e^{-\Delta\sigma \mathcal{H}_0/\epsilon}\right) 
1/\sqrt{2\epsilon}\,[\mathcal{D},x^{M_2} ] \,\chi_{M_2} 
(\sigma_1).
\end{equation}
If we suppose that these insertions are not near any other insertions,
we may assume that there are projectors onto the states of zero energy to 
the left and right.  This allows us to rewrite our expression as
\begin{multline} \label{eq:collisionterm}
  \,x^{M_1} \chi_{M_1}(\sigma_2)
  \frac{1}{2 \epsilon}  \mathcal{D}^2 
\left(e^{-\Delta\sigma \mathcal{H}_0/\epsilon}\right) 
  \,x^{M_2} \,\chi_{M_2} (\sigma_1)
\\ 
= -\,x^{M_1} \chi_{M_1} (\sigma_2)
  \mathcal{H}_0
 \sum_\lambda 
   |\lambda\rangle 
    \frac{e^{-\Delta\sigma E_\lambda /\epsilon}}{\epsilon} 
   \langle \lambda | 
\,x^{M_2} \,\chi_{M_2} (\sigma_1).
\end{multline}
Note, however, that 
\[
\lim_{\epsilon \to 0} \frac{e^{-\Delta\sigma E_\lambda /\epsilon}}{\epsilon} 
  = \frac{1}{E_\lambda} \delta(\Delta \sigma),
\]
unless $E_\lambda = 0$.  Fortunately, the $E_\lambda = 0$ states do
not contribute, as they are killed by the extra factor of
$\mathcal{H}_0$ in equation (\ref{eq:collisionterm}).  

Hence, integrating (\ref{eq:collisionterm}) over $\Delta \sigma$ gives
\begin{multline} \label{eq:collisionterm2}
  -\,x^{M_1} \chi_{M_1}
 \sum_{\lambda \ne 0}
   |\lambda\rangle 
   \langle \lambda | 
\,x^{M_2} \,\chi_{M_2}
 = 
 -\,x^{M_1} \chi_{M_1}
 \bigl( 1 - \sum_{\rho}
   |\rho\rangle 
   \langle \rho | \bigr)
\,x^{M_2} \,\chi_{M_2} (\sigma_1)\\
= x^{M_1} \chi_{M_2}
 \sum_{\rho}
   |\rho\rangle 
   \langle \rho|
\,x^{M_2} \, \chi_{M_1},
\end{multline}
where in the last line we have used the antisymmetry of the two
$\chi_M$ fields. Consider that the term (\ref{eq:collisionterm2})
will always appear sandwiched between projectors $\sum_\rho
|\rho\rangle \langle \rho|$.  Taking a bra from the left projector and
a ket from the right projector, we get the matrix element
\begin{equation}
  \langle \rho_2 | x^{M_1} \chi_{M_2}
 \sum_{\rho}
   |\rho\rangle 
   \langle \rho|
\,x^{M_2} \, \chi_{M_1} |\rho_1 \rangle 
= \mathbf{X}^{M_1}_{\rho_2 \rho}  
  \mathbf{X}^{M_1}_{\rho \rho_1} \chi_{M_2}\chi_{M_1}
= \frac{1}{2} [\mathbf{X}^{M_2}, \mathbf{X}^{M_1}]_{\rho_2 \rho_1}
 \chi_{M_2} \chi_{M_1}.
\end{equation}
The upshot of this computation is that, including the contributions from the $i x^M \pi_M$ terms, our boundary
state becomes
\begin{equation}
  \int \mathcal{D} \pi \, \mathcal{D} \chi \, 
  \text{trP}\exp
    \biggl\{
        i\, \mathbf{X}^M \pi_M 
      + \frac{1}{2}[\mathbf{X}^{M},\mathbf{X}^N] \chi_M \chi_N
    \biggr\}
|\pi^M\rangle |\chi^M \rangle.
\end{equation}
This is precisely the boundary state of a collection of D0-branes
\cite{Asakawa:2003ax,Kraus:2000nj} up to one subtlety which we now
describe.

Since the states $|\rho\rangle$ satisfy $\mathcal{D} |\rho\rangle =
0$, they can be organized into positive and negative chirality states.
Since none of the objects we are considering switch the chirality of
the states, the boundary state is actually a sum of two terms, one
that uses only positive chirality states and one that uses only
negative chirality states.

Up till now, we have been working in the NS sector.  However,
if we consider the same calculation in the RR sector, we need to account
for the change in boundary conditions of the fermions.  This is
accomplished by putting a factor of the $\Gamma$ inside the
over trace.  This has the effect of switching the sign of the
negative chirality states and, hence, their RR charge.
The interpretation is simple: the positive chirality states represent
D0-branes while the negative chirality states represent
D$\bar{\text{0}}$-branes. Note, that this implies that the
D0-brane charge of the brane is given by the index theorem \cite{AS,Eguchi:jx}
\[
  \# \text{D0s}- \# \text{D}\bar{\text{0}}s
 = \text{index}(\mathcal{D})
 = \int \widehat{A} \wedge \text{ch}(F),
\]
where $\widehat{A}$ is the A-roof genus of the brane manifold.  
This correctly reproduces the D0-brane charge formula found in
\cite{Green:1996dd,Cheung:1997az,Minasian:1997mm,Freed:1999vc}, in the case when
the spacetime background is flat.

To summarize, we have found a simple map between a brane configuration
and a matrix configuration of D0 and D$\bar{\text{0}}$-branes.  One
simply finds the zero modes of the Dirac operator, which we
denote $|\rho\rangle$, and then computes the matrix
\[
    \mathbf{X}^M_{\rho_1 \rho_2} 
     = \langle \rho_1 | x^M (\xi) | \rho_2 \rangle.
\]

\section{Derivation of the Born-Infeld action from the matrix action } 
\label{s:BI}

As a concrete check that our map makes sense, we derive the equations
of motion of the abelian-Born-Infeld action using the equations of
motion of the matrices.  This sort of derivation is similar to those
found in
\cite{Ishibashi:1998ni,Seiberg:1999vs,Cornalba:2000wq,Hyakutake:2001kn}
and is essentially a rephrasing in matrix language of the fact that
the form of the Born Infeld action is invariant under the
Seiberg-Witten map \cite{Seiberg:1999vs}.

Recall that the the Born-Infeld (BI) action is derived in the
approximation of slowly varying field strength
\cite{Abouelsaood:1986gd,Fradkin:1985qd}.  Thus, we can think of the
BI action as giving the equations of motion for a small perturbation
of the gauge field around a constant $F_{MN}$ background.  Hence, to
derive the BI action from the matrix action we must calculate the
matrices associated to a constant $F_{MN}$ background plus a small
fluctuation.

The plan of the calculation is as follows.  First, we find
the zero modes of the Dirac operator in a constant $F_{MN}$ background
and use our map to construct the associated matrices.  We then perturb
this background by $\delta A_M$ and calculate the change in the
matrices.  Finally, we expand the matrix equation of motion to lowest
order in $\delta A_M$ and check that the resulting equation is
equivalent to the BI-action equations of motion.
We can also consider small perturbations in the location of the brane.
These turn out to be much more trivial so we will only sketch how they
work out at the end.

\subsection{The matrices for a constant $F_{MN}$ background}

Our first goal is to solve for the zero modes when $F$ is constant.  We will
add on fluctuations later.  Our Hamiltonian is given by
\[
  H = - \frac{1}{2} (\partial_J - i A_J)^2 
      + i \frac{1}{4} \Gamma^J \Gamma^K F_{JK}.
\]
By using rotations we can assume that the only non-vanishing
components of $F_{MN}$ are the $f_i = F_{2i-1\,2i}$.  We chose the gauge
$A_{2i-1} = -f_i \,x^{2i}$.

Furthermore, we take our states to be eigenvectors of
$\Sigma^{2i-1\,2i}=-\frac{i}{4} \Gamma^{2i-1} \Gamma^{2i}$, with
eigenvalue $s_i = \pm \frac{1}{2}$.  It will also be useful to define
$p_i = x_{2i-1}$ and $q_{i} = x_{2i}$. With these conventions, the
eigenstates of $H$ are given by the eigenfunctions of the simple
harmonic oscillator in the even directions and plane waves in the odd
directions.
\begin{equation}
|\Psi_{k,n,s}\rangle=
\prod_i \frac{1}{\sqrt{2\pi}}
   \left(\frac{\sqrt{f_i}}{2^{n_i} (n_i)!\sqrt{\pi}}\right)^{1/2}
  e^{i k_i p_i}    \mathcal{H}_{n_i}
\left(
   \sqrt{f_i}
  \left(q_{i}+\frac{k_i}{f_i}\right)
\right)\exp\left\{-\frac{f_i}{2} \left(q_{i} + \frac{k_i}{f_i}\right)^2\right\}
|s\rangle,
\end{equation}
where the $\mathcal{H}_n$ are the Hermite polynomials.  The energies
are given by $\mathcal{E}_{k,n,s} =\sum_i f_i ( n_i-s_i+\frac{1}{2}) $.  Note
that the $k$, $n$ and $s$ labels without the $i$ subscript are used as
collective indices.  

The lowest energy states have zero energy. They are given by taking $n_i = 0$ and
$s_i = \frac{1}{2}$ and will be denoted $|\Psi_{k}\rangle$.  These
states determine
\begin{align} \label{eq:pi}
  \mathbf{X}_0^{2i-1} &= \langle \Psi_k| p_i |\Psi_{k'}\rangle = 
   -i \partial_{k_i'} \delta(k-k'), \\ \label{eq:qi}
 \mathbf{X}_0^{2i} &= \langle \Psi_k| q_i |\Psi_{k'}\rangle = 
    -\frac{k_i}{f_i}\delta(k-k'),
\end{align}
where $\delta(k-k') = \prod_i \delta(k_i-k_i')$.  This determines the
commutation relation
\[
  [\mathbf{X}_0^{2i-1} , \mathbf{X}_0^{2i} ]
     = -\frac{i}{f_i} \boldsymbol{1}.
\]
Where $\boldsymbol{1}$ is the matrix $\delta(k-k')$.  This is
as we found in the bosonic theory.
Covariantly, we can write this as
\begin{equation}\label{eq:commutator}
  [\mathbf{X}_0^{M} , \mathbf{X}_0^{N} ] = i \,\theta^{MN},
\end{equation}
where $\theta^{MN}$ is the inverse of $F_{MN}$.

\subsection{The shift in the matrices under a shift $\delta A_M$}

We now want to consider small fluctuations, $\delta A_J$, about the
constant $F$ background.  Computing the change in the matrices
under this shift is straightforward, but requires some algebra. 
The reader interested in just
the results can skip to subsection \ref{s:BIderivation} after examining 
equations (\ref{eq:XdX}) and (\ref{eq:adjointaction}).  Consider the Dirac operator for our
background
\[
  \mathcal{D} = \Gamma^J (-i\partial_J - A_J).
\]
Under our perturbation, we get $\mathcal{D} \to \mathcal{D}+\delta
\mathcal{D}$, where
\[
  \delta \mathcal{D} = -\Gamma^J \delta A_J.
\]
Our states are also perturbed: $|\Psi_{k}\rangle \to
|\Psi_{k}\rangle + |\delta \Psi_k\rangle$.  Since we want to find the
new ground states, we must solve
\[
  (\mathcal{D}+\delta \mathcal{D}) 
      (|\Psi_{k}\rangle + |\delta \Psi_k\rangle) = 0,
\]
which to lowest order gives
\begin{equation}
  \mathcal{D} |\delta \Psi_k\rangle = - \delta \mathcal{D} |\Psi_{k}\rangle,
\end{equation}
or, equivalently,
\begin{equation}
 2 H |\delta \Psi_k\rangle = - \mathcal{D} \delta \mathcal{D} |\Psi_{k}\rangle,
\end{equation}
which has the solution
\[
  |\delta \Psi_k\rangle = \sum_{\mathcal{E}_{k',n,s} \ne 0} |\Psi_{k,n,s}\rangle
   \frac{1}{2 \mathcal{E}_{k',n,s}}\langle \Psi_{k,n,s} |
\mathcal{D} \Gamma^J \delta A_J |\Psi_k\rangle.
\]
This is just the usual formula from degenerate perturbation theory.  

We would now like to compute the shift in the matrices, 
$\mathbf{X}_0^M \to \mathbf{X}_0^M +\delta \mathbf{X}^M$, generated
by the shift in the gauge field.  To do this, we first need to evaluate
the quantity
\begin{equation}\label{eq:qexpectation}
  \langle \Psi_{k''} |q_i| \delta\Psi_{k}\rangle =
 \sum_{\mathcal{E}_{k',n,s} \ne 0}\langle \Psi_{k''} |q_i
 |\Psi_{k',n,s}\rangle  \frac{1}{2\mathcal{E}_{k',n,s}}\langle
 \Psi_{k',n,s} | \mathcal{D} \Gamma^J \delta A_J |\Psi_k\rangle.
\end{equation}
Note that the $q_i$ integral in $\langle
\Psi_{k''} |q_i |\Psi_{k',n,s}\rangle$ will fix $k' = k''$.  Moreover,
since $q_i$ has no effect on the spins, we may assume $s^i =
\frac{1}{2}$.  It is useful to rewrite the $q^i$ in terms of raising and
lowering operators,
\begin{equation}\label{eq:raisingformula}
q_i = -\frac{k_i}{f_i} + \frac{a_i+a_i^{\dagger}}{\sqrt{2f_i}},
\end{equation}
where the constant term, $-\frac{k_i}{f_i}$, arises because
the center of our oscillator potential is shifted from the origin.
Using the fact that $a_i |\Psi_k\rangle = 0$, and $\langle
\Psi_{k} |\Psi_{k',n,s}\rangle=0$ for $n \ne 0 $, we see that
\begin{equation}
  \langle \Psi_{k''} |q_i|\Psi_{k',n,s}\rangle 
 = 
  \langle \Psi_{k''} |\frac{a_i}{\sqrt{2 f_i}} |\Psi_{k',n,s}\rangle,
\end{equation}
which implies that we must have $|\Psi_{k',n,s}\rangle = a_i^{\dagger}
|\Psi_{k'}\rangle$ for the matrix element not to vanish, and, hence,
that
\[
  \langle \Psi_{k''} |q_i| \delta\Psi_{k}\rangle = \frac{1}{2 f_i}
 \langle\Psi_{k''} |\frac{a_i}{\sqrt{2 f_i}} \mathcal{D} \Gamma^J \delta
 A_J |\Psi_k\rangle.
\]
Now, using equation (\ref{eq:raisingformula}),
as well as the identities, $\langle \Psi_k | a_i^\dagger = 0$ and
$\langle \Psi_k | \mathcal{D} = 0$ we get
\begin{equation}
  \langle \Psi_{k''} |q_i| \delta\Psi_{k}\rangle =
-\frac{1}{2 f_i}
 \langle\Psi_{k''} | [\mathcal{D},q_i] \Gamma^J \delta
 A_J |\Psi_k\rangle 
 = -\frac{1}{2 f_i}
 \langle\Psi_{k''} | i \Gamma^{2i} \Gamma^J \delta
 A_J |\Psi_k\rangle.
\end{equation}
This expression can be simplified still further:
\begin{equation}
  -\frac{1}{2 f_i}
 \langle\Psi_{k''} | i \Gamma^{2i} \Gamma^J \delta
 A_J |\Psi_k\rangle 
%
=
-\frac{1}{f_i}\frac{i}{4} \langle\Psi_{k''} | 
(4 i \Sigma^{2i,J}+2 \delta_{2i,J}) \delta
 A_J |\Psi_k\rangle,
\end{equation}
where $\Sigma^{MN} = -\frac{i}{4} [\Gamma^{M}, \Gamma^N]$ is our
Lorentz generator.  Note that $\langle\Psi_{k''} | \Sigma^{2i,J}
|\Psi_k\rangle$ vanishes unless $J = 2i-1$, in which case we can
replace $\Sigma^{2i,2i-1}$ with $-\frac{1}{2}$.  Hence we get
\[
  \langle \Psi_{k''} |q_i| \delta\Psi_{k}\rangle = -\frac{1}{2 f_i}
   \langle\Psi_{k''} | \delta A_{2i-1} + i \delta A_{2i}
   |\Psi_k\rangle.
\]
A similar calculation gives
\[
   \langle \Psi_{k''} |p_i| \delta\Psi_{k}\rangle
   = \frac{1}{2 f_i}
   \langle\Psi_{k''} | \delta A_{2i} - i \delta A_{2i-1}
   |\Psi_k\rangle.
\]
Consider that, under the perturbation by $\delta A$, the matrices
$\mathbf{X}_0^J$ are shifted via $\mathbf{X}_0^J \to \mathbf{X}_0^J +\delta
\mathbf{X}^J$, where
\[
  \delta \mathbf{X}^J_{k'k} = 
  \langle \Psi_{k'} | x^J | \delta\Psi_{k}\rangle
  + \langle \delta \Psi_{k'} | x^J | \Psi_{k}\rangle.
\]
Using the explicit form of $\langle \Psi_{k'} | x^J |
\delta\Psi_{k}\rangle $, this reduces to simply
\begin{equation} \label{eq:XdX}
 \delta \mathbf{X}^J_{k'k} = 
    \langle \Psi_{k'}| \theta^{JK} \delta A_{K} |\Psi_{k}\rangle.
\end{equation}
Note the similarity of this form of the shift in $\mathbf{X}^J$ to the
standard
expressions~\cite{Ishibashi:1998ni,Seiberg:1999vs,Cornalba:2000wq,miao1996,Hyakutake:2001kn}.

\subsection{One final identity}

Before we can derive the Born-Infeld action from the matrix action,
we need one more identity.
We need to understand the action, by commutator, of the
$\mathbf{X}_0^J$ on  a matrix of the form 
$\mathbf{G}_{k,k'} = \langle \Psi_k | g(x) | \Psi_k'\rangle$,
where $g(x)$ is an arbitrary function.  We have from equations
(\ref{eq:pi}) and (\ref{eq:qi}) that
\begin{align}
  [ \mathbf{X}_0^{2i},\mathbf{G}]_{k',k}
 &= \frac{1}{f_i} (k_i-k'_i)  \left(\mathbf{G}\right)_{k',k},\\
  [ \mathbf{X}_0^{2i-1},\mathbf{G}]_{k',k}&=
  i (\partial_{k_i'}+\partial_{k_i}) \left(\mathbf{G}\right)_{k',k}.
\end{align}
Now consider,
\begin{align}
  \frac{1}{f_i} (k_i-k'_i) & \langle\Psi_{k'} | g(x^I) |\Psi_k\rangle\nonumber \\
 &= \frac{1}{f_i}\int dx\,  \,
   \left((k_i -k'_i) \,\Psi_{k'}^* \Psi_k\, g(x^I)\right)\nonumber \\
 &=\frac{1}{f_i} \int dx \, \bigl(-i\partial_{p_i} (\Psi_{k'}^*\Psi_k)\bigr)
   g(x^I)\nonumber \\
&=\frac{1}{f_i} \int dx \,  \Psi_{k'}^*\Psi_k
   \bigl(i\partial_{p_i}g(x^I)\bigr)  \nonumber  \\
&=\frac{1}{f_i} \langle\Psi_{k'} | (i \partial_{p_i} g(x^I)) |\Psi_k\rangle.
\end{align}
Similarly, we have
\begin{align}
  i (\partial_{k_i'}+\partial_{k_i}) \langle\Psi_{k'} | g(x^I) |\Psi_k\rangle
 = \frac{1}{f_i} \langle\Psi_{k'} |(-i\partial_{q_i} g(x^I)) |\Psi_k\rangle.
\end{align}
Taken together, these two identities tell us that we can replace the
adjoint action of $\mathbf{X}^J_0$ with $i\theta^{JK} \partial_K$.  
Specifically, we have
\begin{equation} \label{eq:adjointaction}
   [\mathbf{X}^J_0 , \langle \Psi_k | f(x) | \Psi_k'\rangle ]  = 
   \langle \Psi_k| i\theta^{JK} \partial_K f(x) |\Psi_k'\rangle.
\end{equation}
Again, the form of this equation is expected, since the commutator is
approximately given by the Poisson bracket at large field 
strength~\cite{Cornalba:2000wq}. This gives us all the identities we need.

\subsection{Derivation of the Born-Infeld action} \label{s:BIderivation}

Before we derive the full Born Infeld action, we consider first just
the lowest order terms in $1/F^2$.  These come from the
lowest order term in the matrix action, given by
\[
   \frac{1}{4} \text{tr} [\mathbf{X}_M,\mathbf{X}_N]^2,
\]
whose equations of motion are
\[
   [\mathbf{X}^I, [\mathbf{X}_I,\mathbf{X}^J]]=0.
\]
Consider expanding the matrix equations of motion
around $\mathbf{X}^I_0$ to first order  in $\delta \mathbf{X}^I$,
\begin{multline}
  [\mathbf{X}^I,[\mathbf{X}_I,\mathbf{X}^J]] = 
  [\mathbf{X}^I_0,[{\mathbf{X}_0}_I,\mathbf{X}^J_0]]
+ [\delta \mathbf{X}^I,[{\mathbf{X}_0}_I,\mathbf{X}^J_0]]\\
+ [ \mathbf{X}^I_0,[\delta \mathbf{X}_I,\mathbf{X}^J_0]]
+ [ \mathbf{X}^I_0,[{\mathbf{X}_0}_I,\delta \mathbf{X}^J]]
+\mathcal{O}(\delta \mathbf{X}^2).
\end{multline}
Using the fact that
$[\mathbf{X}^I_0,\mathbf{X}^J_0] 
\propto \boldsymbol{1}$, we get the lowest order condition
\[
  [ \mathbf{X}^I_0,[{\mathbf{X}_0}_I,\delta \mathbf{X}^J]]
- [ \mathbf{X}^I_0,[\mathbf{X}^J_0,\delta \mathbf{X}_I]] = 0.
\]
Applying our formulas, (\ref{eq:XdX}) and (\ref{eq:adjointaction}), this
equation reduces to
\[
  \langle \Psi_k | \bigl(-{\theta^{J}}_M \theta^{MK} \partial_J F_{KI} \bigr)
\theta^{MJ} |\Psi_{k'}\rangle = 0,
\]
which can only vanish if
\[
    {\theta^{J}}_M \theta^{MK} \partial_J F_{KI} = 0.
\]
As expected, this is nothing but the lowest order term in the $1/F$
expansion of the Born-Infeld equations of motion.  This computation
is similar to the one found in~\cite{Cornalba:2000wq}.

We now seek an all orders derivation.  For this we require some
knowledge of the terms in the matrix action to all powers of
$\mathbf{X}^I$.  The action is given by the dimensional reduction of
the non-abelian Born Infeld action to $0+0$ dimensions.
Unfortunately, the full non-abelian Born Infeld is not known beyond
order $\alpha'^4$ \cite{Koerber:2002zb}.  However, it is known that if
one ignores terms with covariant derivatives and commutators of field
strengths, then the full action is given by the symmetrized trace
prescription found in \cite{Tseytlin:1997cs}.

Fortunately, it is easy to see that such terms cannot contribute in
our approximation.  Consider that, under dimensional reduction, we
have $F_{MN}\to -i(2\pi\alpha')^{-2}[\mathbf{X}_N,\mathbf{X}_M]$.  To
zeroth order then, $F_{MN} \to (2\pi\alpha')^{-2} \theta_{MN}
\boldsymbol{1}$, where we have used equation (\ref{eq:commutator}).
Since we are expanding to lowest order, at most one $F$ is not
proportional to the identity.  Hence, in a term with a commutator of
two $F$s, one of the two $F$s must be proportional to the identity and
the commutator must vanish.

A similar argument shows that terms with covariant derivatives cannot
contribute.  Under dimensional reduction, a covariant derivative is
replaced by a commutator.  Note that we must always have at least two
derivatives and, using integration by parts, we can assume that the
derivatives act on different $F$s.  Then, as we argued before, one of
the two $F$s must be proportional to the identity and will vanish when
acted on by a commutator.

The upshot of this is that we may use as our all orders action 
\cite{Tseytlin:1997cs}
\[
 \text{STr}\,  \sqrt{\det \left(\delta_{MN} \boldsymbol{1} -
  i (2\pi\alpha')^{-1}[\mathbf{X}_M,\mathbf{X}_N]\right)},
\]
where $\text{STr}$ stands for symmetrized trace and the determinant
acts on the $M,N$ indices.  In this prescription one formally expands
out the action in powers of the commutator and then sums with equal
weight over each ordering of the commutators.

The equations of motion for this action are given by
\begin{equation}\label{eq:matrixeom}
  [\mathbf{X}^I , \text{S}\biggl\{
    \sqrt{\det \left(\delta_{MN} \boldsymbol{1} -
   i (2\pi\alpha')^{-1}[\mathbf{X}_M,\mathbf{X}_N]\right)}  
    \left(\frac{C}{1-C^2}\right)_{IJ}
  \biggr\} ] = 0,
\end{equation}
where by $\text{S}$ we mean symmetrization over all commutators and we
have defined the matrix $C_{MN} = -i
(2\pi\alpha')^{-1}[\mathbf{X}_M,\mathbf{X}_N]$. Note that this matrix
is multiplied both through its spacetime indices and through ordinary
matrix multiplication.  Expanding equation (\ref{eq:matrixeom}) to
first order in $\delta \mathbf{X}^K$ and using equations
(\ref{eq:XdX}) and (\ref{eq:adjointaction}), we find
\begin{equation}
  \sqrt{\det \left(\delta^{MN}  +  
(2\pi\alpha')^{-1} \theta^{MN}\right)}  
 \times   \left(\frac{\tilde{\theta}^2}{1-\tilde{\theta}^2}\right)_{MN} 
\partial^M F^{NK} 
\left(\frac{\tilde{\theta}}{1-\tilde{\theta}^2}\right)_{KJ}
    =0,
\end{equation}
where $\tilde{\theta} = (2\pi\alpha')^{-1} \theta$.
This is equivalent to the condition
\begin{equation}
  \left(\frac{\tilde{\theta}^2}{1-\tilde{\theta}^2}\right)_{MN} \partial^M F^{NK} 
  =  \left(\frac{1}{1-(2\pi\alpha'F)^2}\right)_{MN} \partial^M F^{NK}   = 0.
\end{equation}
These are just the familiar equations of motion of the Born-Infeld action.

It is straightforward, although somewhat trivial, to extend these
results to the full Dirac-Born-Infeld action by include transverse
fluctuations of the brane.  If $X^A$ is a transverse direction, and we
consider a shift $\delta X^A$, then the change in the matrices, to
lowest order, is simply given by $\delta \mathbf{X}^A_{k'k} = \langle
\Psi_{k'}| \delta X^A| \Psi_{k} \rangle$.  In other words, to lowest
order we can consider the zero modes unchanged.  Substituting this
shift in the matrices back into the matrix equations of motion gives
the expected equations of motion for the transverse fluctuations.

It would be interesting to try to extend these results to the next
order in $\delta A_M$.  One could use the same procedure we have
used here, but attempt to find the second order shift in the
$\mathbf{X}^I$.  This requires doing second order perturbation theory
to find the new corrected zero modes.  One could then attempt to use
the matrix action to compute derivative corrections to the Born-Infeld
action and compare with the results of \cite{Wyllard:2000qe}.

\section{Decoupling of the higher modes of the endpoint} \label{s:decoupling}

In this section we would like to understand in detail how the higher
modes of the endpoint decouple.  To do this we consider the case of a
flat D2-brane with uniform flux.  This setup is convenient, as the
associated world sheet theory is exactly solvable
\cite{Abouelsaood:1986gd} and the dual D0-brane description takes a
simple form. Moreover, as mentioned in the introduction, in this case
there is a proof given by N.~Ishibashi \cite{Ishibashi:1998ni} that
the two setups are equivalent.

As we will see below, the proof \cite{Ishibashi:1998ni} of the
equivalence of the D0-brane and D2-brane descriptions is essentially a
proof that one may take the endpoint of the string to live in the
ground state. Unfortunately, the proof given in
\cite{Ishibashi:1998ni} contains various singularities which we would
like to avoid.  Moreover, we would like to study the problem in a
framework where we can unambiguously isolate the behavior of the
endpoint of the string which cannot be done in the continuum framework
considered in \cite{Ishibashi:1998ni}.  This motivates us to
reconsider the argument of \cite{Ishibashi:1998ni} in a latticized
model of the string world sheet action.

In a latticized model, with lattice spacing $\epsilon$, the endpoint at $\sigma = 0$
becomes a charged particle with mass $\epsilon$ that is coupled to its
neighboring lattice site by a spring-like force with spring constant
$1/\epsilon$. It is convenient to take the other endpoint at $\sigma = \pi$ 
to live on a brane without any magnetic flux, since otherwise
the string spectrum would be independent of the flux
\cite{Abouelsaood:1986gd}.

If we ignore the coupling of the endpoint to its neighbor for the
moment, the action of the endpoint is just that of the familiar Landau
problem.  Upon quantization, we find a ground state labeled by a
momentum $k$ and an infinite tower of evenly spaced excited states
separated by an energy $f/\epsilon$ where $f$ is the magnitude of the
background magnetic field.
Notice that as $\epsilon$ is taken to zero, the gap between the energy
levels of the endpoints goes to infinity.  This does not imply,
however, that the endpoint is constrained to the ground state, as the
force between the endpoint and its neighbor also goes to infinity in
this limit.  

To determine whether or not the endpoint leaves the ground state we
perform the following computation, the details of which are given in
appendix \ref{s:lattice}: Construct the operators $A$ and $A^\dagger$
which raise and lower the string from one energy level to another.
The operator $N = A^\dagger A$ is then the operator which gives the
Landau level.  We can then attempt to calculate the expectation value
of N.  As an example, consider $\langle 0 | N |0\rangle $, where
$|0\rangle$ is the ground state of the string, and take the limit as
$\epsilon$ goes to zero.  This calculation is not easy to perform
analytically, but, in the large $f$ limit it is given by
\[
  \langle 0 | N |0\rangle = \frac{4}{3\pi} \,\, \frac{1}{f^3}
  +\mathcal{O}(\frac{1}{f^4}).
\]
The important point here is that the average value of $N$ is non-zero.
We do see that, as $f$ goes to infinity, the endpoint becomes
constrained to the ground state as expected, but, away from this
limit, there is a definite sense in which the endpoint is free to move
to higher energy levels.

We would now like to study how these excitations of the endpoint decouple 
in the continuum limit.  
It is easy to check that, for any two finite norm
states, $\psi_1$ and $\psi_2$, the inner product, $\langle \psi_1| A |
\psi_2 \rangle$, goes to zero as $\epsilon$ goes to zero.  Indeed, the
non-zero contribution of $\langle 0 | N |0\rangle $ comes entirely
from the normal ordering of $A^\dagger A$ in terms of the oscillators
of the string.  This implies that, in fact, for finite norm states
$\psi$, the expectation value $\langle \psi | N |\psi\rangle $ is
independent of the choice of $\psi$ in the continuum limit.

What these results are saying is that physics below the cutoff scale
is insensitive to whether or not the endpoint of the string is in the
ground state.  These results can be made even more definitive.
As in \cite{Ishibashi:1998ni}, define a new Hamiltonian, $H_0$, in which
the endpoint is always constrained to its ground state.  This
Hamiltonian is easily constructed by sending the mass of the endpoint
to zero in the Lagrangian and then canonically quantizing.  It is the
straightforward to show that
\[
   \lim_{\epsilon\to 0}  (H - H_0) = 0.
\]
In other words, the theory in which the string endpoint is constrained
to the ground state and the theory in which it is allowed to leave the
ground state are completely equivalent in the small $\epsilon$ limit.
Note that this limit is only really valid when the expression is 
sandwiched between to finite norm fock space states.
This decoupling of the endpoint excitations is the basic mechanism
behind the duality between the D2-brane and the D0-brane descriptions.
Indeed, the Hamiltonian $H_0$ is precisely the Hamiltonian of a string
ending on an infinite collection of D0-branes with non-commutative
coordinates $\mathbf{X}$ and $\mathbf{Y}$ satisfying
$[\mathbf{X},\mathbf{Y}] = i/f$.

There is one aspect of this result which at first seems paradoxical. 
Consider that, in the full quantization of
the string there is an oscillator $b_0$, which, for small $f$, raises
the string from one Landau level to another.  This seems to contradict
the fact that the endpoint of the string can be assumed to be in the
ground state.

To understand how this can happen, consider the following toy model
similar to the one considered in \cite{Bigatti:1999iz}.  Take a
charged particle which is constrained to it's lowest energy level so
that its coordinates obey $[x,y] = i/f$.  Take a second particle of
zero charge and mass $m$ and couple it to the charged particle using a
spring with spring constant $k$.  The Hamiltonian of this model is the
same as the projected Hamiltonian of the string $H_0$ in the case when
there are only two lattice sites.  The theory has two zero modes $p$
and $q$ which satisfy $[p,q] = i/f$, and two oscillators $a$ and $b$.
The Hamiltonian is given by
\[
   H  = \mathcal{E}_+ a^\dagger a+ \mathcal{E}_- b^\dagger b,
\]
where
\[
   \mathcal{E}_\pm = \sqrt{\frac{k}{m}} \sqrt{1+\frac{mk}{2f^2}
   \left(1\pm \sqrt{1+\frac{4f^2}{mk} }\right)}.
\]
Consider what happens when $f$ is very small.  In this case
$\mathcal{E}_+$ becomes very large so that the $a$ oscillator
decouples.  On the other hand $\mathcal{E}_-$ is given, to lowest order,
by $f/m$.  Thus, the Hamiltonian is approximately given by $H = f/m
b^\dagger b$.  This Hamiltonian is nothing but the ordinary Landau
Hamiltonian for a particle with mass $m$.  In other words, when we
couple a particle in the lowest Landau level to a massive particle,
the massive particle behaves, in the weak field limit, as though it were
a massive charged particle.  Thus, the fact that the string can be
raised to higher Landau levels is not inconsistent with the fact that
the endpoint is trapped in the lowest Landau level.

\section{Conclusions} \label{s:conclusions}

We have now discussed some general features of smooth and fuzzy
Dp-branes and constructed a map from one to the other.  Moreover,
we have checked in a few elementary examples that this map gives 
the expected results.

Unfortunately,
we have had very little to say about smooth branes.
The basic problem with understanding smooth branes is coming up with a
consistent framework in which they can be studied.  Indeed, at this
stage we have only really defined them as branes which cannot be
described as matrix valued configurations of D0-branes.

The example of the D2-brane with a single unit of flux should provide
a natural laboratory for studying such branes, since the fuzzy
description is completely inapplicable.  In principle, one can try to
manipulate the boundary state for such a brane in the same way as we
did for the fuzzy case, but use a different regulator in which the
$\epsilon \to 0$ limit, and the limit as the cutoff is removed are
simultaneous.  One can then try to expand the resulting boundary state
around the small size limit to try and understand the physics of a
single D0-brane expanding into a small D2-brane.

We have attempted this computation for a few choices of such a
regulator and the results imply that one finds something like the
boundary state of a D0-brane plus extra terms which are suppressed by
factors of $e^{-F}$ where $F$ is the field strength on the D2-brane.
Note that this non-perturbative suppression in the large $F$ limit is
required, as there is no mode living on a single D0-brane which allows
it to expand into a D2-brane in perturbation theory.  These results
are suggestive, but one also finds various new divergences which need
to be subtracted, and more information is required to extract any
sensible physics from the computation.

Another subject which we have not discussed is whether other branes
besides the flat brane with uniform flux have an exact duality between
smooth and fuzzy descriptions.  Here it seems likely that the BPS nature
of the flat brane is important.  
It is not clear, however, how this works out in more general examples.

\section*{Acknowledgments}
I would like to thank W.\ Taylor for getting me interested in this
problem, for useful comments on a draft and for numerous
discussions.  I would also like to thank A. Hashimoto, Y.\ He and M.\
Van Raamsdonk for their comments on a draft.  I would like to
thank K. Hashimoto and M. Van Raamsdonk for sharing some of their
thoughts on fuzzy D2-branes and I. Singer for helping me understand
aspects of the index theorem. Finally, I would like to thank S.\
Robinson, J.\ Shelton, and B.\ Zwiebach for useful discussions. This
work was supported in part by the DOE through contract
\#DE-FC02-94ER40818 and through funds from the University of
Wisconsin.

\appendix

\section{Details of the lattice model} \label{s:lattice}

In this appendix, we give a few details of the lattice model discussed
in section \ref{s:decoupling}.  We consider our brane to have only two
coordinates $X^1$ and $X^2$.  We take the string to have $n+1$ lattice
sites with fields labeled $X_m$, where $i$ runs from $0$ to $n$.  We
take $X_0$ to be the endpoint of a string living on a brane with
magnetic flux.  The other endpoint $X_n$ is taken to have Neumann
boundary conditions.  The lattice spacing, $\epsilon$ is given by
$\pi/n$.

We take as our lattice model
\begin{equation}
 \mathcal{L} = \sum_{m = 0}^{n} \frac{1}{2} \epsilon \dot{X}_m^2
-\sum_{m = 1}^n \frac{1}{2 \epsilon} (X_m - X_{m-1})^2
-\frac{1}{2} f(X_0^1 \dot{X}_0^2 - X_0^2 \dot{X}_0^1).
\end{equation}
The model is solved in essentially the same way as the continuum model
\cite{Abouelsaood:1986gd}.  First we define the complex coordinates
$X^{\pm} = (X^1\pm i X^2)/\sqrt{2}$.  We can then expand in terms of
modes that solve the classical equations of motion,
\[
  X^+_m = x^+ + i \sum_{k>0} a_k \psi_k(m) - i \sum_{k<0}
  b^\dagger_{-k} \psi_k(m),
\]
where 
\[
  \psi^+_k(m) = \mathcal{N} \cos(\pi k m /n - \pi k (1+1/2n))e^{i
  \mathcal{E}_k \tau}.
\]
The energies, $\mathcal{E}_k$, are given by the dispersion
relation
\[
  \mathcal{E}_k = \frac{2}{\epsilon} \sin(\epsilon k/2),
\]
and the momenta, $k$, are required to satisfy
\begin{equation}\label{eq:dispersion1}
    f = - \frac{\sin(\pi k (1+1/n))}{\cos(\pi k (1+1/2n))},
\end{equation}
for $-n\le k \le n$.

The normalization, $\mathcal{N}$, is determined by requiring
$[a_k,a_k^\dagger] = 1$ and $[b_k,b_k^\dagger] = 1$.  The full
expression is quite complex, however, at large $n$, it is
approximately given by $\mathcal{N} = \sqrt{|\mathcal{E}_k/\pi|}$.

We can now study the physics of the endpoint $X_0$.  If we decouple
the rest of the string, the endpoint Lagrangian is just the usual
Lagrangian for a charged particle with mass $\epsilon$ in the presence
of a magnetic field.  Diagonalizing the endpoint Hamiltonian just
gives the usual Landau levels.

The raising and lowering operators that take us from one Landau level
to another are given by $A = \epsilon \dot{X}_0^- /\sqrt{f} $ and
$A^\dagger = \epsilon \dot{X}_0^+/\sqrt{f}$, which satisfy
$[A,A^\dagger] = 1$.  Note that since $\dot{X}_0$ is finite as
$\epsilon \to 0$, both $A$ and $A^\dagger$ go to zero as a formal
expansion of oscillators.

On the other hand, consider defining the number operator $N =
A^\dagger A$.  If we denote normal ordering by $::$, we can write $N =
:A^\dagger A: + c_0$.  where $c_0$ is a constant.  The operator
$:A^\dagger A:$ now simply goes to zero as $\epsilon \to 0$ when it is
sandwiched between any two finite norm fock space states, but the
constant $c_0$ does not.  We have been unable to achieve an analytic
expression for $c_0$, although a numerical estimate is quite easy to
compute.  However, if we assume $f$ to be large, it is
straightforward to expand the relation \ref{eq:dispersion1} for large
$f$ and sum the series analytically.  This gives
\[
  c_0 =  \frac{4}{3\pi f^3} + \mathcal{O} (1/f^4),
\]
as stated in section \ref{s:decoupling}.  In this sense, the one finds that the
endpoint does leave the ground state.

Now we reproduce the argument of \cite{Ishibashi:1998ni} by showing
that in the $\epsilon \to 0$ limit the endpoint can be assumed to be
in the ground state.  The easiest way to put the endpoint into the
ground state is just to remove its kinetic term.  The only affect this
has on our construction is to change the relation \ref{eq:dispersion1}
to
\begin{equation} \label{eq:dispersion2}
  f = - \frac{\sin(\pi k )}{\cos(\pi k(1+1/2n))}.
\end{equation}
At finite $n$, this has a definite effect on the energy spectrum of the
theory.  As $n\to \infty$, however, both relations
\ref{eq:dispersion1} and \ref{eq:dispersion2} become simply $f = -
\tan{\pi k}$, so that there is no effect on the spectrum.  As a final
note, a reader familiar with \cite{Ishibashi:1998ni}, may wonder what
happened to the counter terms that were introduced in that version of
the argument.  In the latticized model they appear in the expansion of
coupling of the endpoint to its neighboring lattice site and are,
thus, naturally included.


\end{document}